\pdfoutput=1

\documentclass[fleqn,usenatbib]{mnras}

\usepackage{newtxtext,newtxmath}

\usepackage[T1]{fontenc}

\DeclareRobustCommand{\VAN}[3]{#2}
\let\VANthebibliography\thebibliography
\def\thebibliography{\DeclareRobustCommand{\VAN}[3]{##3}\VANthebibliography}

\usepackage{graphicx}	
\usepackage{amsmath}	
\usepackage{caption}
\usepackage{subcaption}

\newcommand{\microns}{\,$\upmu\mathrm{m}$ } 

\title[Strongly lensed submm galaxies in future surveys]{
A large population of strongly lensed faint submillimetre galaxies in future dark energy surveys inferred from {\it JWST} imaging}

\author[J. Pearson et al.]{James Pearson$^{1}$\thanks{E-mail: james.pearson@open.ac.uk},
Stephen Serjeant$^{1}$,
Wei-Hao Wang$^{2}$,
Zhen-Kai Gao$^{2}$,
Arif Babul$^{3}$, 
Scott Chapman$^{4}$, 
\newauthor 
Chian-Chou Chen$^{2}$, 
David L. Clements$^{5}$, 
Christopher J. Conselice$^{6}$, 
James Dunlop$^{7}$,
Lulu Fan$^{8}$,
Luis C. Ho$^{9}$,
\newauthor
Ho Seong Hwang$^{10,11}$,
Maciej Koprowski$^{12}$, 
Micha\l\ J. Micha{\l}owski$^{13}$, and
Hyunjin Shim$^{14}$
\\
$^{1}$School of Physical Sciences, The Open University, Milton Keynes, MK7 6AA, UK\\
$^{2}$Academia Sinica Institute of Astronomy and Astrophysics (ASIAA), No. 1, Sec. 4, Roosevelt Rd., Taipei 10617, Taiwan\\
$^{3}$Department of Physics \& Astronomy, University of Victoria, 3800 Finnerty Road, Victoria, BC V8P 5C2, Canada\\
$^{4}$Department of Physics \& Astronomy, Vancouver Campus, University of British Columbia, 325 - 6224 Agricultural Road, Vancouver, BC V6T 1Z1, Canada\\
$^{5}$Astrophysics Group, Imperial College London, Blackett Laboratory, Prince Consort Road, London SW7 2AZ, UK\\
$^{6}$Department of Physics and Astronomy, The University of Manchester, Alan Turing Building, Oxford Road, Manchester, M13 9PL, UK\\
$^{7}$Institute for Astronomy, Royal Observatory, Blackford Hill, Edinburgh EH9 3HJ, UK\\
$^{8}$Deep Space Exploration Laboratory / Department of Astronomy, University of Science and Technology of China, Hefei 230026, China\\
$^{9}$Kavli Institute for Astronomy and Astrophysics, Peking University, 5 Yi He Yuan Road, Haidian District, Beijing 100871, P. R. China\\
$^{10}$Astronomy Program, Department of Physics and Astronomy, Seoul National University, 1 Gwanak-ro, Gwanak-gu, Seoul 08826, Republic of Korea\\
$^{11}$SNU Astronomy Research Center, Seoul National University, 1 Gwanak-ro, Gwanak-gu, Seoul 08826, Republic of Korea\\
$^{12}$Institute of Astronomy, Faculty of Physics, Astronomy and Informatics, Nicolaus Copernicus University, Grudziądzka 5, 87-100 Toru\'{n}, Poland\\
$^{13}$Astronomical Observatory Institute, Faculty of Physics, Adam Mickiewicz University in Pozna\'{n}, ul. S\l oneczna 36, 60-286 Pozna\'{n}, Poland\\
$^{14}$Department of Earth Science Education, Kyungpook National University, 80 Daehak-ro, Buk-gu, Daegu 41566, Republic of Korea\\
}

\date{Accepted 2023 December 14. Received 2023 November 28; in original form 2023 September 4}

\pubyear{2023}

\begin{document}
\label{firstpage}
\pagerange{\pageref{firstpage}--\pageref{lastpage}}
\maketitle

\begin{abstract}

Bright galaxies at sub-millimetre wavelengths from {\it Herschel} are now well known to be predominantly strongly gravitationally lensed. The same models that successfully predicted this strongly lensed population also predict about one percent of faint 450\microns-selected galaxies from deep James Clerk Maxwell Telescope (JCMT) surveys will also be strongly lensed. Follow-up ALMA campaigns have so far found one potential lens candidate, but without clear compelling evidence e.g. from lensing arcs. 
Here we report the discovery of a compelling gravitational lens system confirming the lensing population predictions, with a $z_{\rm s}=3.4\pm0.4$ submm source lensed by a $z_{\rm spec}=0.360$ foreground galaxy within the COSMOS field, identified through public {\it JWST} imaging of a 450\microns source in the SCUBA-2 Ultra Deep Imaging EAO Survey (STUDIES) catalogue.
These systems will typically be well within the detectable range of future wide-field surveys such as {\it Euclid} and {\it Roman}, and since sub-millimetre galaxies are predominantly very red at optical/near-infrared wavelengths, they will tend to appear in near-infrared channels only. Extrapolating to the {\it Euclid}-Wide survey, we predict tens of thousands of strongly lensed near-infrared galaxies. This will be transformative for the study of dusty star-forming galaxies at cosmic noon, but will be a contaminant population in searches for strongly lensed ultra-high-redshift galaxies in {\it Euclid} and {\it Roman}.

\end{abstract}








\section{Introduction}

Strong galaxy-galaxy gravitational lensing is the phenomenon in which light from a background galaxy (source) is bent around a foreground galaxy (lens) due to gravity, resulting in a distorted view of the source reaching an observer, typically in the form of extended arcs, rings, and multiple images of the source. Observations of these objects can provide a wealth of information, from testing general relativity and dark matter models by constraining the mass profile of the foreground galaxy dark matter halo \citep{1995ApJ...449...18M,2015ApJ...800...94S,2018Sci...360.1342C} and detecting substructures within the halo and along the line of sight \citep{2007ApJ...668..806M,2014MNRAS.442.2017V,2016ApJ...823...37H,2022MNRAS.510.2464A,2022MNRAS.511.3046H,2022ApJ...927...83O,2023ApJ...942...75W}, to helping understand galaxy formation and evolution by studying the inner mass profile slope of the lens \citep{2017SCPMA..60h0411C,2017MNRAS.471.3079H,2021A&A...651A..18S} and the properties of high-redshift background sources \citep{2010Natur.464..733S,2011MNRAS.410.1687D,2015MNRAS.452.2258D,2018ApJ...866L..12G,2018ApJ...864L..22S,2020MNRAS.497.1654C,2021ApJ...906..107K}. Specific types of strong lenses can also aid in constraining cosmological parameters, with the matter density $\Omega_{\rm m}$ and dark energy equation of state parameters $w_{0}$ and $w_{\rm a}$ constrained by double source plane lenses \citep{2012MNRAS.424.2864C,2021MNRAS.505.2136S,2022JCAP...07..033S}, while time delays between different images of the same source can put constraints on the Hubble constant $H_{0}$ from lensed quasars \citep{2010ARA&A..48...87T,2021MNRAS.501..784D,2021MNRAS.504.2224L} and supernovae \citep{2019PhRvL.123w1101C,2021ApJ...910...65B,2021A&A...653A..29B}.

These applications require large statistical samples of lenses. Following earlier work on the fraction of lensed submillimetre galaxies (SMGs) in blank field SMG surveys \citep{2002MNRAS.335L..17C}, an early breakthrough of the {\it Herschel} space observatory was the observation that nearly all the bright SMGs (500\microns flux densities above ${\sim}100$\,mJy) are strongly gravitationally lensed, once obvious contaminant populations of blazars and very nearby galaxies are removed \citep{2010Sci...330..800N}. This enabled a strong lens selection method with an extraordinary ${\sim}100$\% efficiency. Further campaigns at longer wavelengths with data from the South Pole Telescope found a similarly efficient strong lens selection \citep{2020ApJ...902...78R}. Since SMGs are typically very red at optical and near-infrared wavelengths \citep[e.g.][]{2016ApJ...820...82C,2017MNRAS.469..492M,2020ApJ...889...80L,2022MNRAS.514.2915S}, follow-up redshift campaigns have mostly relied on CO and [CI] transitions, yielding several hundred secure redshifts to date \citep[e.g.][]{2020A&A...635A...7N,2020ApJ...902...78R,2022MNRAS.511.3017U}, which in turn have enabled a very wide range of galaxy evolution studies exploiting the strong gravitational magnification \citep[e.g.][]{2018NatAs...2...56Z,2023MNRAS.522.2995B,2023MNRAS.521.5508H,2023ApJ...948...44R}. 

The search for strong gravitational lenses is on the cusp of another transformative change. Upcoming surveys such as \textit{Euclid} \citep{2011arXiv1110.3193L} and the Vera Rubin Legacy Survey of Space and Time \citep[LSST;][]{2019ApJ...873..111I} are set to discover around $10^{5}$ of them amongst a larger population of ${\sim}10^{9}$ galaxies in the optical and near-infrared, with near-infrared imaging of approximately 17000 strong lenses \citep{2020RNAAS...4..190W} also expected by the future \textit{Nancy Grace Roman Space Telescope} \citep{2012arXiv1208.4012G,2015arXiv150303757S,2019arXiv190205569A}.

Most of the effort to date in making lensing predictions for these future dark energy surveys, and in designing machine learning algorithms to identify the lensed systems, has focused on the higher angular resolution visible wavelength imaging \citep[e.g.][]{2011arXiv1110.3193L,2015ApJ...811...20C,2019A&A...625A.119M}. In this work, we show that the relatively neglected near-infrared channels will also be a very rich source of strong gravitational lenses. We do this by using new public {\it JWST} imaging of the 450\microns SCUBA-2 SMGs in the COSMOS field to search for strong gravitational lensing features, demonstrating that the faint SMG population has a very high surface density of strong lensing systems visible in the near-infrared.

This paper is organised as follows. The catalogue of SCUBA-2 SMGs in the COSMOS field and the corresponding {\it JWST} data are discussed in Section \ref{sec:data}. The results of the investigation are presented in Section \ref{sec:results}, including photometry and modelling of the identified gravitational lens and its impact on the inferred number density of strongly-lensed SMGs. This is followed by a discussion of our findings in Section \ref{sec:discussion} and final conclusions in Section \ref{sec:conclusions}.

\section{Data}
\label{sec:data}



\subsection{Submm Observations}

The 450\microns imaging used in this work was taken by the Submillimetre Common-User Bolometer Array 2 \citep[SCUBA-2;][]{2013MNRAS.430.2545C,2013MNRAS.430.2534D,2013MNRAS.430.2513H} on the James Clerk Maxwell Telescope (JCMT) as part of the SCUBA-2 Ultra Deep Imaging EAO Survey \citep[STUDIES;][]{2017ApJ...850...37W} within the 2\,square degree Cosmological Evolution Survey \citep[COSMOS;][]{2007ApJS..172....1S} field. Centred on (ra, dec) coordinates (10:00:28.0, +02:24:00) \citep{2013MNRAS.436.1919C, 2020ApJ...889...80L}, STUDIES observations covered 450 square arcminutes down to a depth of 0.59 mJy per beam at 450\microns with the resulting data set containing 479 identified SMGs (Gao et al., in preparation).

\subsection{Infrared Observations}



The {\it JWST} data consisted of publicly available Near Infrared Camera (NIRCam) and Mid-Infrared Instrument (MIRI) observations of the COSMOS field as part of the COSMOS-Web \citep{2023ApJ...954...31C} and Public Release IMaging for Extragalactic Research \citep[PRIMER;][]{2021jwst.prop.1837D} surveys. 
Within the COSMOS field, PRIMER in part aims to cover 144 square arcmin in 8 NIRCam filters 
and 112 square arcmin in two MIRI filters. 
Likewise, COSMOS-Web aims to cover 0.54 square degrees in four NIRCam filters 
down to 5$\sigma$ depths of ${\sim}28$ magnitudes, and 0.19 square degrees in one MIRI filter 
down to 5$\sigma$ depths of ${\sim}26$ magnitudes.

Data was acquired through the Space Telescope Science Institute (STScI) Mikulski Archive for Space Telescopes (MAST) Portal, covering the STUDIES area using a search radius of 15 arcmin centred on the STUDIES central coordinates following \cite{2020ApJ...889...80L}. As the COSMOS-Web and PRIMER surveys were not yet completed, only 59 per cent of the search area had been surveyed at the time of this work. All available COSMOS-Web and PRIMER calibration level 3 imaging data within the search area was acquired: COSMOS-Web data was partially available for all filters (NIRCam F115W, F150W, F277W, and F444W along with MIRI F770W); PRIMER data was partially available for eight out of ten filters (NIRCam F115W, F200W, F277W, F356W, F410M, and F444W along with MIRI F1800W and F770W; missing NIRCam F090W and F150W).

\section{Results}
\label{sec:results}

\subsection{Photometry}
\label{subsec:photometry}

\begin{figure}
    \centering
    \begin{subfigure}[b]{0.495\columnwidth}
        \centering
        \includegraphics[width=\textwidth]{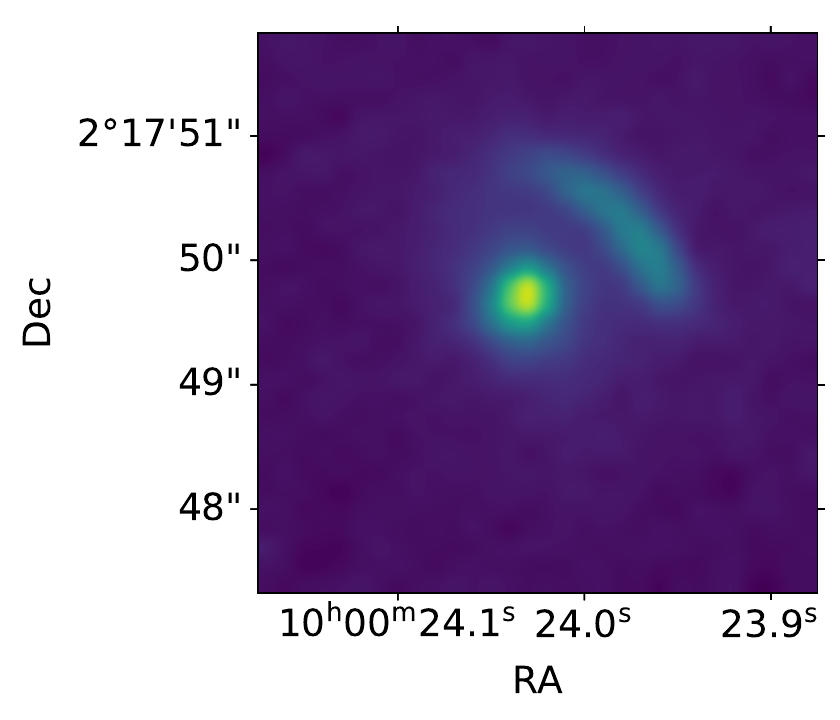}
    \end{subfigure}
    \hfill
    \begin{subfigure}[b]{0.495\columnwidth}
        \centering
        \includegraphics[width=\textwidth]{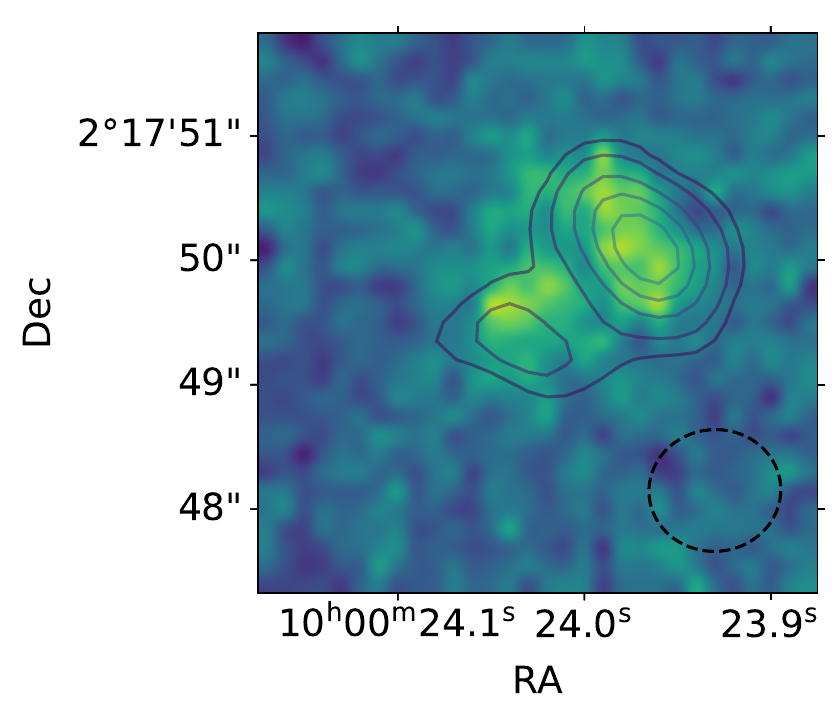}
    \end{subfigure}
    \caption{{\it JWST} MIRI F770W (left) and F1800W (right) imaging of the strong gravitational lens from PRIMER as $4.5\times4.5$ arcsec cutouts centred on the visible foreground lens. The lensed background source lies just to its upper right, clearly visible as an arc in the left image. ALMA imaging is presented as overlaid contours on the right image, depicting the lensed background source as a bright image and fainter counterimage, with no foreground lens visible. The ALMA beam FWHM is shown as the dashed line in the right image. Cutouts have been reprojected to align their sky orientation with figure axes, inline with ALMA observations.}
    \label{fig:jwst_lens_f770w_f1800w}
\end{figure}

At present, out of the 479 SMGs identified at 450\microns by STUDIES, 283 were found to have coordinates within regions imaged by {\it JWST} in at least one filter: 63 SMGs were imaged in COSMOS-Web and 263 in PRIMER. From visually inspecting the corresponding {\it JWST} NIRCam and MIRI imaging, a single strong lens candidate was discovered, whose measured properties from STUDIES are given in Table\,\ref{tab:studies_lens}. The foreground lens of this lensing system corresponded to the known galaxy WISEA J100023.99+021749.6, at (ra, dec) coordinates (10:00:24.025, +02:17:49.54) with a spectroscopic redshift of $0.35978\pm0.00018$ \citep{2018ApJS..234...21D}, which has not as yet been classed as a gravitational lens in previous studies. The SMG itself has submm colours clearly strongly inconsistent with the foreground lens redshift, as the 450\microns:850\microns (deboosted) flux density ratio of ${\sim}1.9$ (Table\,\ref{tab:studies_lens}) requires a redshift of $z>3$ \citep[e.g.][]{2020ApJ...889...80L}, indicative of a background source. Imaging at the SMG's position had only been taken in {\it JWST} MIRI filters F770W and F1800W, with the central foreground lens and lensed background source seen in both, as shown in Fig.\,\ref{fig:jwst_lens_f770w_f1800w}. The arc of the lensed background source is clearly visible in the higher resolution F770W filter, and two separate objects are still resolved in the F1800W filter despite its lower resolution.

\begin{table*}
	\centering
	\caption{STUDIES SCUBA-2 data for the 450\microns SMG. Columns are as follows: object name; right ascension (J2000) (deg); declination (J2000) (deg); signal-to-noise ratio; raw flux density (mJy); instrumental noise at the 450\microns SMG position (mJy); deboosted flux density (mJy); total noise (consisting of instrumental noise, flux-deboosting uncertainty, and confusion noise) at 450\microns (mJy); deboosted flux density of the SMG at 850\microns (mJy); and the total noise at 850\microns (mJy).}
	\label{tab:studies_lens}
        \setlength{\tabcolsep}{3pt}
	\begin{tabular}{lccccccccr}
		\hline
		ID & RA (deg) & Dec (deg) & SNR & $S_{450\rm, raw}$  & $\delta S_{450\rm, raw}$  & $S_{450\rm, total}$  & $\delta S_{450\rm, total}$  & $S_{850\rm, total}$ & $\delta S_{850\rm, total}$ \\
		\hline
            STUDIES-COSMOS-450-019 /  & 150.100117 & 2.297500 & 15.79 & 19.2 & 1.2 & 18.5 & 2.3 & 9.76 & 0.65\\ 
            STUDIES-COSMOS-850-006 &  &  &  & mJy & mJy & mJy & mJy & mJy & mJy\\ 
  \hline
	\end{tabular}
\end{table*}

For cross-matching, the background source in PRIMER imaging is located at 10:00:23.98 +02:17:50.26, while the STUDIES 450\microns source is located 1.054 arcsec away at 10:00:24.03 +02:17:51.00. The point-spread function (PSF) of the SCUBA-2 beam has a full width at half maximum (FWHM) of 7.9 arcsec for 450\microns \citep{2013MNRAS.430.2534D,2020ApJ...889...80L}, hence neighbouring galaxies near the lensing system could not be ruled out as candidates for the SMG from this data alone. 
However, publicly available high-resolution Atacama Large Millimeter/submillimeter Array (ALMA) submm observations from the ALMA public archive confirmed the lensed background source to be the SMG. This source was found to be the bright source AS2COS0005.1 
observed in the 870\microns AS2COSMOS catalogue of the brightest SCUBA-2 SMGs in COSMOS \citep{2019ApJ...880...43S,2020MNRAS.495.3409S}, with a likely counter image detected close to the foreground lens given by the nearby source AS2COS0005.2, which is $4.0\pm0.7$ times fainter than the main extended image.
These 870\microns ALMA observations of AS2COS0005.1/5.2 are shown here as contours overlaying the PRIMER {\it JWST} F1800W image in Fig.\,\ref{fig:jwst_lens_f770w_f1800w}.

PRIMER {\it JWST} photometry is given in Table\,\ref{tab:jwst_primer_lens} for the foreground lens and background source separately. As there existed no {\it JWST} F1800W photometry for the lensing system, the F1800W flux densities were calculated based on the same lens and source segmentation applied to the F770W photometry, with uncertainties based on the 1$\sigma$ background. Photometry at other wavelengths exists for only a single object: that of the lens, the source, or the combined lensing system, depending on the wavelength. Table\,\ref{tab:ned_photometry} contains the publicly available flux densities of the lensing system in other wavelengths or passbands, with those shorter than 5\microns and those longer than 100\microns corresponding to the lens and the source, respectively.

To obtain a photometric redshift of the lensed source, the 450\microns and 850\microns flux densities in Table\,\ref{tab:studies_lens} along with the {\it Herschel} SPIRE and MAMBO photometry in Table\,\ref{tab:ned_photometry} were used to fit a spectral energy distribution (SED) using the ``Eyelash'' template, which was used to fit the $z>2$ SMG of the Cosmic Eyelash gravitational lens \citep[see e.g.,][]{2010Natur.464..733S,2017MNRAS.467..330I,2018MNRAS.477.1099D}, resulting in a redshift of $z_\mathrm{source}=3.41\pm0.14$. However, SMGs exhibit a range of intrinsic colour temperatures \citep[e.g.][]{Blain+03}, resulting in larger photometric redshift uncertainties \citep[e.g.][]{Casey+20}. Recently, \citet{2023MNRAS.522.2995B} compared spectroscopic redshifts in bright lensed {\it Herschel} galaxies with photo-$z$ from the Eyelash template, using submm and mm-wave data from ALMA and SPIRE; the diversity of SEDs in that sample implies an irreducible uncertainty in our photo-$z$ of $\pm0.4$, so we adopt $z_\mathrm{source}=3.4\pm0.4$. 

\begin{table*}
	\centering
	\caption{PRIMER {\it JWST} MIRI F770W and F1800W data for the foreground lens and background source. Columns are as follows: object; right ascension (J2000) (deg); declination (J2000) (deg); F770W and F1800W flux densities (mJy); F770W and F1800W AB magnitudes; and F770W ellipticity and position angle counter-clockwise from north (deg). F1800W flux densities were not available from {\it JWST} photometry, so were calculated based on the same lens and source segmentation used for F770W photometry, with uncertainties based on the 1-$\sigma$ background.}
	\label{tab:jwst_primer_lens}
	\begin{tabular}{lcccccccr}
		\hline
		Object & RA (deg) & Dec (deg) & $F_{\nu, F770W}$ (mJy) & $F_{\nu, F1800W}$ (mJy) & $m_{AB, F770W}$ & $m_{AB, F1800W}$ & $e$ & $\phi$ (deg) \\
		\hline
            lens & 150.100134 & 2.297140 & (17.59$\pm$0.07)$\times10^{-3}$ & (6.06$\pm$2.52)$\times10^{-3}$ & 20.787$\pm$0.004 & $21.95^{+0.58}_{-0.38}$ & 0.082 & -17.06 \\
            source & 150.099914 & 2.297295 & (9.81$\pm$0.06)$\times10^{-3}$ & (7.87$\pm$2.77)$\times10^{-3}$ & 21.421$\pm$0.007 & $21.66^{+0.47}_{-0.33}$ & 0.675 & 42.98 \\
  \hline
	\end{tabular}
\end{table*}

\begin{table}
	\centering
	\caption{Combined photometry for the lensing system, excluding {\it JWST} and SCUBA-2 data, presented in Fig.\,\ref{fig:fitted_seds}. 
    Source references: (1) \citet{2008ApJS..175..297A} (2) \citet{2015ApJ...805..121D} (3) \citet{2013wise.rept....1C} (4) \citet{2015ApJS..218...33A} (5) \citet{2021yCat.2368....0S} (6) \citet{2012ApJ...761..140C} (7) \citet{2020MNRAS.495.3409S} (8) \citet{2007ApJS..172..132B}. 
    SDSS data  are obtained via best-fitting de Vaucouleurs and exponential light models. 
    VISTA $K_{s}$ uncertainty is based on the $5\sigma$ depth of the UltraVISTA near-infrared imaging survey \citep{2012A&A...544A.156M}. {\it Spitzer} IRAC data are corrected to total flux from a 2.4" aperture.}
	\label{tab:ned_photometry}
	\begin{tabular}{lcl}
		\hline
		Wavelength / Observed Passband & $F_{\nu}$ (mJy) & Ref.\\ 
		\hline
            4858 \r{A} (SDSS g)	& 0.0039$\pm$0.0005 & (1)\\	
            6290 \r{A} (SDSS r)	& 0.0191$\pm$0.0009 & (1)\\	
            7706 \r{A} (SDSS i)	& 0.0309$\pm$0.0014	& (1)\\ 
            9222 \r{A} (SDSS z)	& 0.0446$\pm$0.0061 & (1)\\ 
            2.2\microns (VISTA $K_{s}$) & 0.121$\pm$0.001	& (2)\\ 
            3.4\microns (WISE W1) & 0.068$\pm$0.006	& (3)\\ 
            3.6\microns ({\it Spitzer} IRAC) & 0.062$\pm$0.004 & (4)\\ 
            4.5\microns ({\it Spitzer} IRAC) & 0.056$\pm$0.004 & (4)\\ 
            4.6\microns (WISE W2) & 0.048$\pm$0.011  & (3)\\ 
            12.0\microns (WISE W3) & $<0.302$ & (3)\\ 
            22.0\microns (WISE W4) & $<2.98$  & (3)\\ 
            24.0\microns ({\it Spitzer} MIPS)    & $<0.0328$ & (5)\\ 
            250\microns ({\it Herschel} SPIRE) & 16.5$\pm$2.6	& (6)\\ 
            350\microns ({\it Herschel} SPIRE) & 25.5$\pm$4.5	& (6)\\ 
            500\microns ({\it Herschel} SPIRE) & 22.6$\pm$6.4	& (6)\\ 
            870\microns (ALMA)     & $8.4^{+0.4}_{-0.3}$  & (7)\\ 
            1.20 mm (MAMBO)        & 5.2$\pm$0.9  & (8)\\ 
  \hline
	\end{tabular}
\end{table}

\begin{figure}
	\includegraphics[width=\columnwidth]{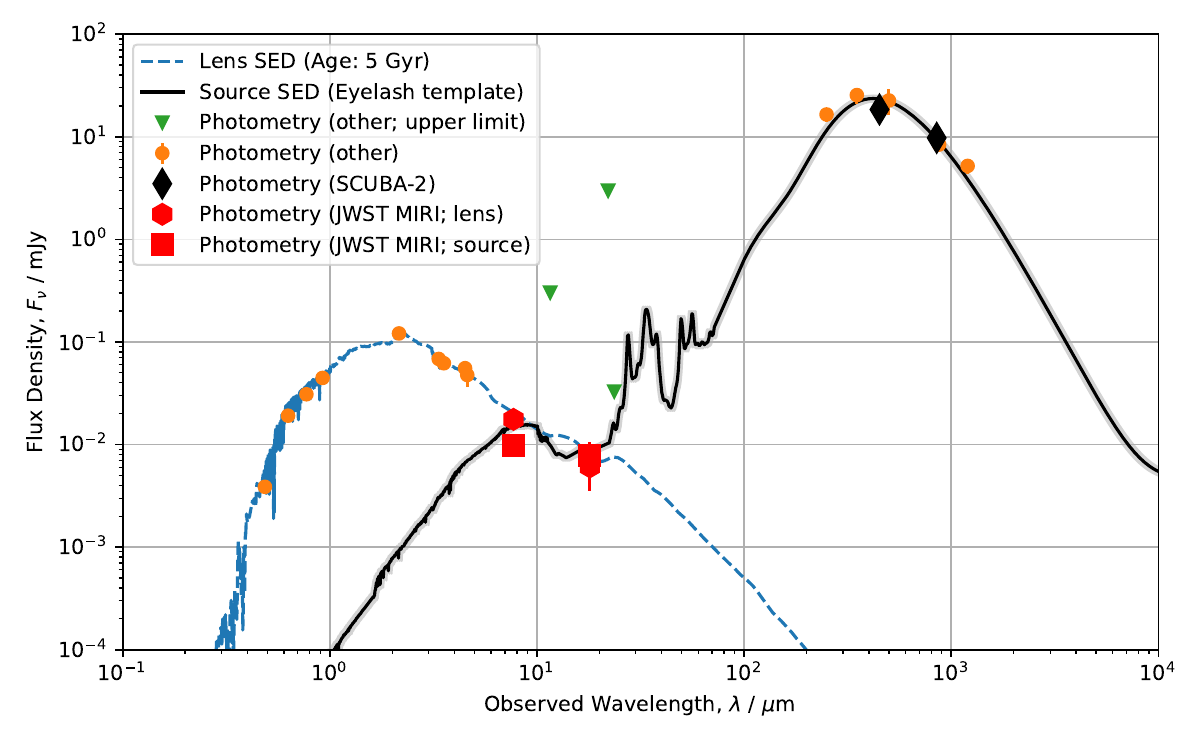}
    \caption{Spectral energy distributions (SEDs) of the foreground lens (blue dashed line) and background source (black solid line), with the shaded region around the latter giving the 1$\sigma$ errors. The lens SED is of a 5 Gyr-old dustless early-type galaxy described in the text at the observed spectroscopic redshift of $z_\mathrm{lens}=0.35978$, fitted to photometry from other surveys (orange circles; see Table\,\ref{tab:ned_photometry}) primarily associated with the lens (those with wavelengths shorter than ${\sim}5$\microns). The source SED is the ``Eyelash'' template used to fit the Cosmic Eyelash gravitational lens, which gave a photometric redshift of $z_\mathrm{source}=3.4\pm0.4$ when fitted to the STUDIES SCUBA-2 photometry (black diamonds; see Table\,\ref{tab:studies_lens}) as well as the {\it Herschel} SPIRE and MAMBO photometry. Also displayed are upper limits of photometry from other surveys (green downward triangles; see Table\,\ref{tab:ned_photometry}) as well as the {\it JWST} MIRI F770W and F1800W photometry for the lens (red hexagon) and source (red square) presented in Table\,\ref{tab:jwst_primer_lens}.}
    \label{fig:fitted_seds}
\end{figure}

To verify that the photometry indicated a combination of lens and source, the flux densities in Table\,\ref{tab:ned_photometry} up to wavelengths of 4.6\microns were then used to fit an SED for the lens at its much lower redshift, taken to be a 5 Gyr-old dustless early-type galaxy \citep[initial mass function from][metallicity $\log_{10}(Z/Z_\odot)=0.0$]{2003PASP..115..763C}, using the Flexible Stellar Population Synthesis for Python package \textsc{pythonFSPS} \citep{2009ApJ...699..486C,2010ApJ...712..833C}\footnote{\tt https://doi.org/10.5281/zenodo.591505}. The resulting SEDs and photometry are presented in Fig.\,\ref{fig:fitted_seds}, which provides a good fit to the overlaid {\it JWST} PRIMER photometry. 
To explore the robustness of the lens SED fitting, a more flexible SED modelling was also performed, fitting a single stellar population with free star formation history, dust extinction, and metallicity. This provided an almost identical fit, and resulted in the following parameters for the foreground lens: $\log_{10}({\rm age/yr})=9.8$, extinction $A_{V}=0$, star formation rate $\log_{10}({\rm SFR})=-1.43$, and stellar mass $\log_{10}({\rm M_{*}/M_\odot})=10.69$, which are consistent with an early-type galaxy. The stellar mass also agrees with the value obtained for the foreground lens from the hCOSMOS redshift survey \citep{2018ApJS..234...21D}, which used the MMT Hectospec multifibre spectrograph to obtain a velocity dispersion $\sigma = 153\pm25\ {\rm km s^{-1}}$ and stellar mass $\log_{10}({\rm M_{*}/M_\odot})=10.55^{+0.14}_{-0.11}$.


While a limit could not be placed on the stellar mass of the background source due to the effects of dust reddening, the star formation rate of the background source could be estimated, for which we used the latest \textit{Planck} cosmology \citep{aghanim2020planck}. Integrating the fitted Eyelash SED presented in Fig.\,\ref{fig:fitted_seds} over the mid and far infrared (8-1000\micron) produced a luminosity of $L_{\rm FIR} = 6.8\times10^{12}\mu^{-1} L_{\odot}$, where $\mu$ is the magnification factor. Using the conversion ${\rm SFR}(M_{\odot}\ {\rm yr^{-1}})=4.5\times10^{-44} L_{\rm FIR}{\rm (ergs\ s^{-1})}$ from \citet{1998ARA&A..36..189K}, results in a star formation rate of ${\sim}1200\ \mu^{-1} M_{\odot}\ {\rm yr^{-1}}$ for a Salpeter initial mass function (IMF). The inferred SFR would be a factor $\times1.5$ lower for a Kroupa IMF \citep[see][also for other caveats on this calculation]{Hayward+14}.

\begin{figure*}
    \centering
    \begin{subfigure}[b]{0.33\textwidth}
        \centering
        \includegraphics[width=\textwidth]{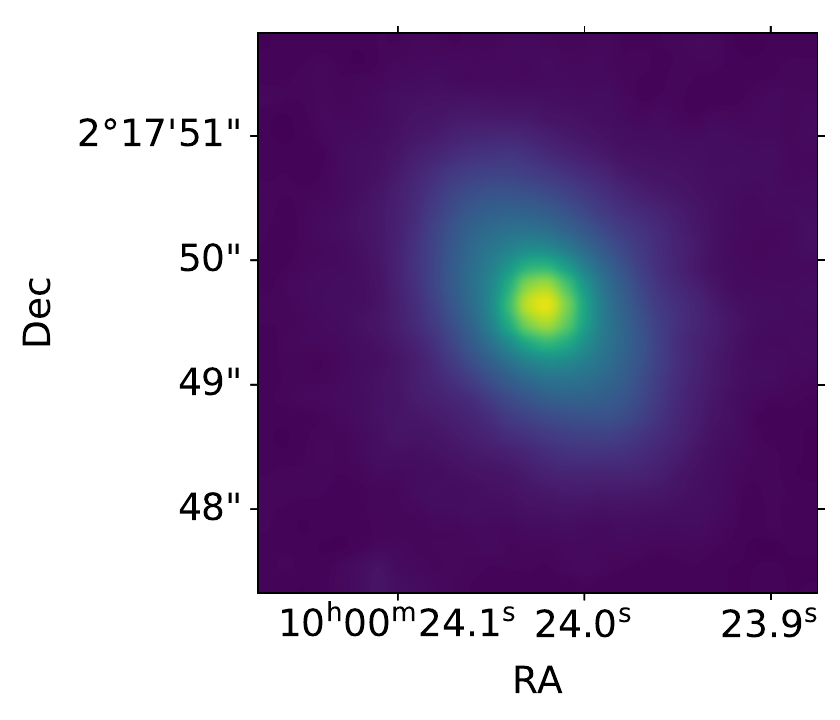}
    \end{subfigure}
    \begin{subfigure}[b]{0.33\textwidth}
        \centering
        \includegraphics[width=\textwidth]{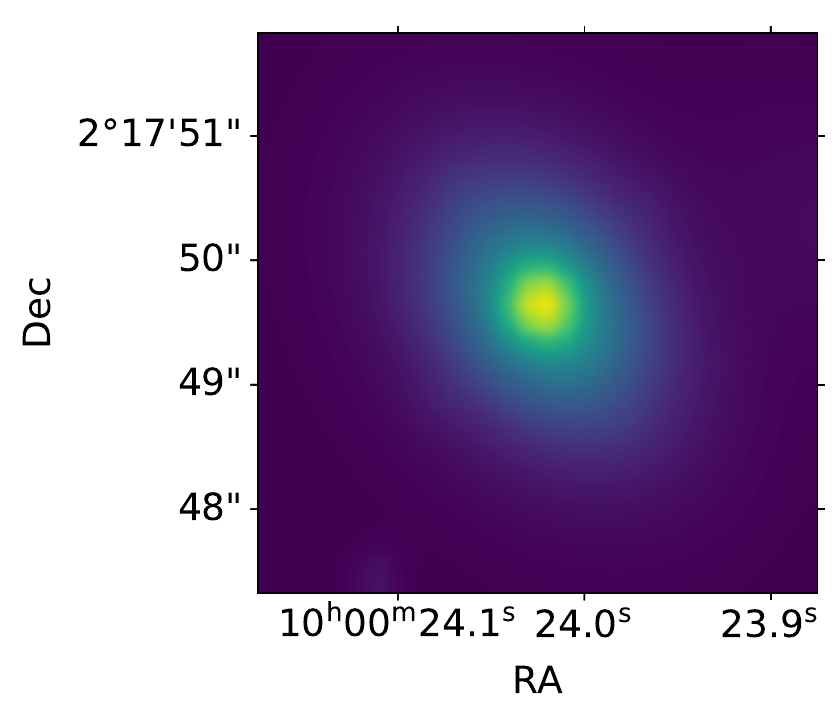}
    \end{subfigure}
    \begin{subfigure}[b]{0.33\textwidth}
        \centering
        \includegraphics[width=\textwidth]{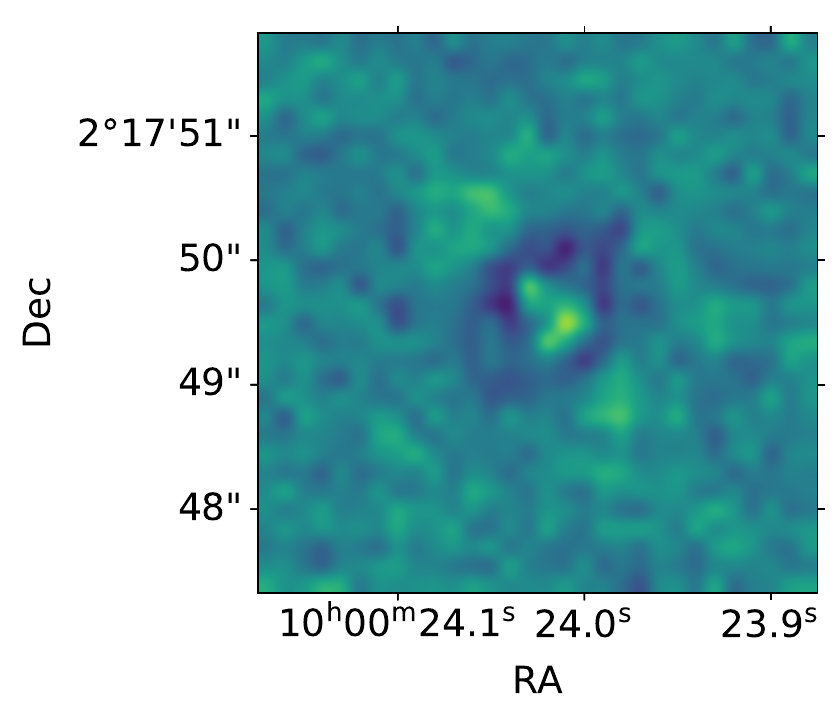}
    \end{subfigure}
    \caption{{\tt GALFIT} lens light modelling and subtraction for publicly-available \textit{Hubble Space Telescope (HST)} F160W near-infrared imaging. The lens light was fitted with a S\'ersic component and an exponential disk component convolved with a F160W model PSF. From left to right: the observed cutout image; the {\tt GALFIT} model for the lens light; and the residual image. All cutouts share the same angular scale and sky orientation matching those of Fig.\,\ref{fig:jwst_lens_f770w_f1800w}, with the first and second likewise presented with logarithmic scaling.}
    \label{fig:hst_modelling}
\end{figure*}

Public {\it Hubble Space Telescope (HST)} WFC3 imaging of the lens was also available from the Cosmic Assembly Near-IR Deep Extragalactic Legacy Survey (CANDELS). However, while the foreground lens was clearly visible across all available bands, there was no obvious background source at these wavelengths. For example, modelling and subtraction of the foreground lens light profile for the near-infrared F160W image of the system was performed using {\tt GALFIT}, fitting the lens light with a S\'ersic component and an exponential disk component convolved with the \textit{HST} F160W model PSF, with no visible background source detected down to a 2$\sigma$ AB magnitude depth of 25.5, as shown in Fig.\,\ref{fig:hst_modelling}.
For comparison, \textit{Euclid} has an average $5\sigma$ limiting AB magnitude of 24.5 in the near-infrared \citep{2022A&A...662A.112E}, so for this lens it would also be unable to detect the background source at this wavelength, but the SMG is both unusually faint in the optical and at unusually high redshift for this flux density (see Section \ref{sec:discussion}). 

\subsection{Lens Modelling}
\label{subsec:lens-modelling}

\begin{figure*}
	\includegraphics[width=\linewidth]{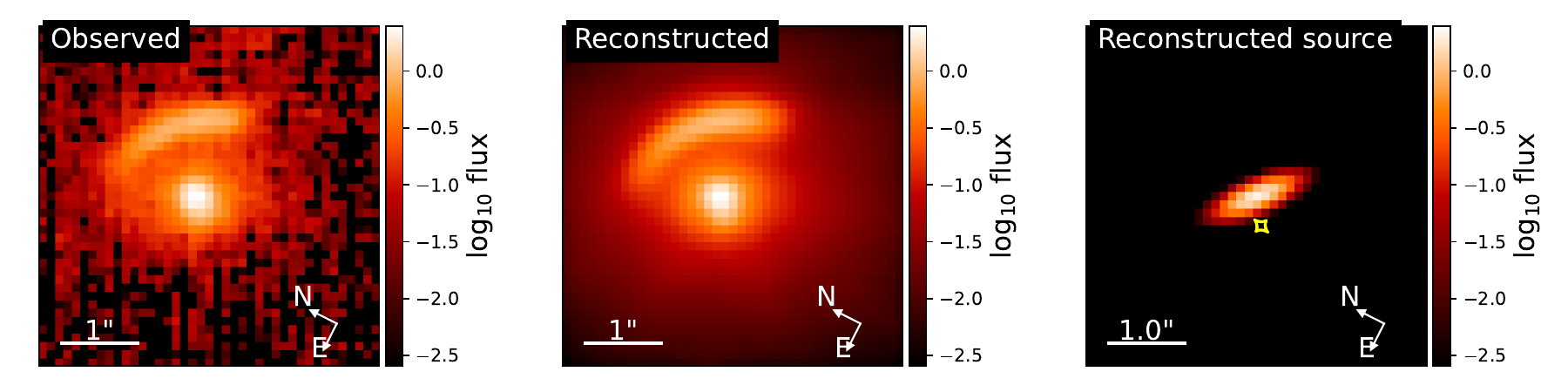}
    \caption{Results from lens modelling. 
    From left to right: the observed cutout image, with original \textit{JWST} sky orientation used for the modelling; the reconstructed image from modelling; and the reconstructed source image centred on the source with the lens model caustic overlaid in yellow. All cutouts share the same angular scale and sky orientation, and are presented with logarithmic scaling.}
    \label{fig:modelling}
\end{figure*}

\begin{table*}
	\centering
	\caption{Modelling parameters of the lens mass profile fitted to the {\it JWST} F770W image, which is centred on the (ra, dec) coordinates (10:00:24.025, +02:17:49.54) of WISEA J100023.99+021749.6. The lens mass was fitted to a singular isothermal ellipsoid (SIE) model with the values given below for each parameter: Einstein radius (arcsec); the two components of complex ellipticity $e_{1}$ and $e_{2}$; the corresponding axis ratio $q$ and position angle $\phi$ (deg) clockwise from east; the central RA \& Dec offset of the mass profile relative to the cutout centre (arcsec); and the total magnification of the lensed source $\mu$. The light profiles of both the lens and source were modelled with elliptical S\'ersic profiles.}
	\label{tab:modelling_results}
	\begin{tabular}{lccccccr}
		\hline
		$\theta_{E}$ (arcsec) & $e_{1}$ & $e_{2}$ & $q$ & $\phi$ (deg) & RA offset (arcsec) & Dec offset (arcsec) & $\mu$ \\
		\hline
            $0.63^{+0.01}_{-0.01}$ & $-0.11^{+0.04}_{-0.04}$ & $-0.07^{+0.03}_{-0.03}$ & $0.77^{+0.08}_{-0.07}$ & $106^{+11}_{-9}$ & $0.10^{+0.01}_{-0.01}$ & $0.15^{+0.01}_{-0.01}$ & $3.0^{+0.3}_{-0.3}$ \\
  \hline
	\end{tabular}
\end{table*}

Lens modelling was performed on the F770W image with \textsc{lenstronomy}\footnote{\tt https://github.com/lenstronomy/lenstronomy} \citep{2018PDU....22..189B,2021JOSS....6.3283B} using a singular isothermal ellipsoid (SIE) lens mass model, along with elliptical S\'ersic profiles for the light models of the lens and source convolved with the \textit{JWST} model PSF for MIRI F770W. Lens modelling was also performed for an elliptical power law (EPL) mass profile as well as modelling with and without external shear, but these offered no better fit than an SIE alone. Due to limited resolution, the mass models produced very similar images despite converging on different parameters, indicating a degeneracy in the modelling. However, the use of a power law or external shear resulted in higher chi-squared errors and excessively elliptical lens mass profiles compared to the lower-ellipticity lens light, indicating that the SIE alone provided the best fit.

Fig.\,\ref{fig:modelling} presents the reconstructed lens and source, the residuals between the observed and modelled lens, and the convergence and magnification maps from the modelling, with the resulting SIE lens mass model parameters given in Table\,\ref{tab:modelling_results}. We follow the definition of ellipticity widely used in lens modelling: for axis ratio $q=b/a$, where $a$ and $b$ are the major and minor axes respectively, and position angle $\phi$ of the major axis, complex ellipticity components are given by
\begin{equation}
    e_{1}= \frac{1-q}{1+q}\cos{2\phi}, \\
    e_{2}= \frac{1-q}{1+q}\sin{2\phi}
	\label{eq:complex_ellipticity}
\end{equation}
such that the parameter space ($e_1$, $e_2$) can be used for more effective fitting.

The modelled lens light has an effective radius $R_{\rm e, lens}=0.61\pm0.03$ arcsec, a S\'ersic index $n_{\rm lens}=5.0^{+0.1}_{-0.2}$ consistent with an early-type galaxy, and an axis ratio $q_{\rm light}=0.73\pm0.02$ closely matching its modelled mass profile. The resulting fit reconstructs the source image with an effective radius $R_{\rm e, source}=0.17\pm0.01$ arcsec, and a low S\'ersic index $n_{\rm source}=0.50^{+0.02}_{-0.01}$ consistent with a late-type galaxy. The unlensed source is also highly elliptical, with an axis ratio $q_{\rm source} = 0.34\pm0.02$, which matches the general results of previous samples of SMGs \citep[e.g.][]{2022ApJ...939L...7C,2022ApJ...936L..19C}. The source's major axis aligns with the direction of elongation of the lensed arc, resulting in a lower magnification than expected: from the ratio of the lensed and unlensed model source flux above the 1$\sigma$ background, a magnification of $\mu=3.0\pm0.3$ is required to convert the highly elliptical source into the observed extended arc. The modelling also uncovered a faint point-source counterimage within a region of the lens light similar to the counterimage observed with ALMA.

Lens modelling was also performed on the ALMA data in order to constrain the SMG source properties at longer wavelengths. For this, the lens model was constrained to the parameters in Table\,\ref{tab:modelling_results} of the {\it JWST} F770W lens model, while the source parameters were left unconstrained and initialised at those predicted by the F770W source model. A magnification of $\mu = 3.0^{+1.5}_{-1.2}$ was required to produce the ALMA source's lensed image and counterimage, in agreement with the magnification obtained from modelling the F770W data, with the modelling limited by comparatively low resolution and a large restoring beam FWHM. However, the modelled (ra, dec) position of the unlensed source was found to be offset by $(0.03\pm0.03, -0.11\pm0.02)$ arcsec relative to that of the source at F770W. This suggests that two different regions of the SMG are responsible for the emission seen in {\it JWST} F770W and ALMA. Such a $\sim$\,kpc-scale offset has precedents in other strongly gravitationally lensed SMGs \citep[e.g.][]{2015MNRAS.452.2258D}.



\begin{figure*}
    \includegraphics[width=15cm]{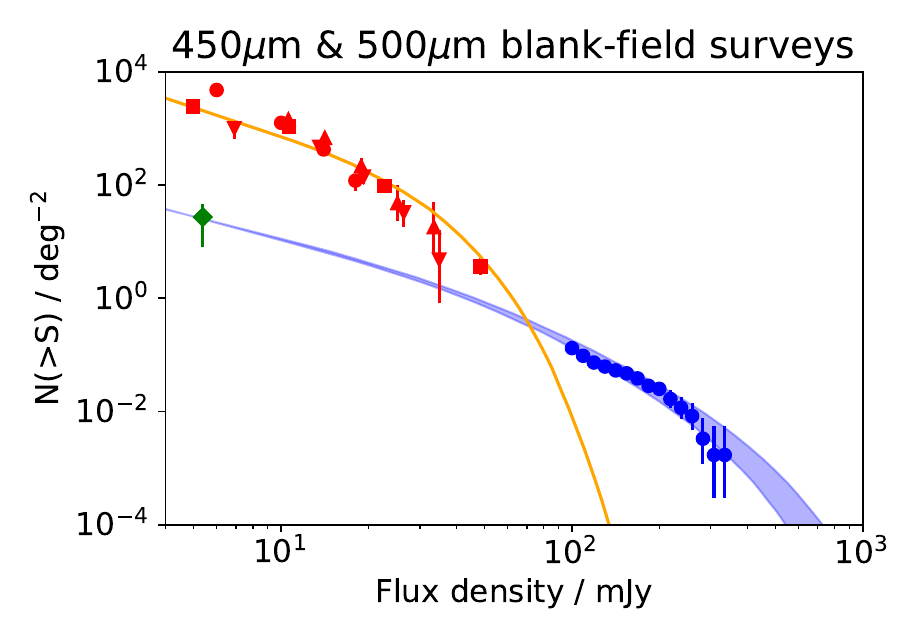}
    \caption{Integral blank-field galaxy number counts at 450-500 \micron. The submm-bright strongly gravitationally lensed population of \citet[][500 \micron]{2017MNRAS.465.3558N} is shown as dark blue filled circles. 
    Our inferred 450\microns surface density of submm-faint strongly lensed galaxies is shown as a dark green diamond, calculated based on the $2\pm\sqrt{2}$ lenses found within the $59\%$ of the 450 square arcminute effective area of STUDIES that has been mapped by {\it JWST} at the time of this work. For this, we use the median 450\microns flux density of the STUDIES catalogue (Gao et al., in preparation) that was searched to obtain the lens candidates.
    The unlensed 450\micron-selected populations are shown in red, as follows: \citet{2017MNRAS.464.3369Z}, red circles; \citet{2013ApJ...776..131C}, downward-pointing red triangles; \citet{2013MNRAS.436.1919C}, upward-pointing red triangles. The unlensed 500\micron-selected population of \citet{2010MNRAS.409..109G} is also shown as red squares.
    The lensed and unlensed population models  are shown as blue and orange curves, respectively \citep[][Negrello, priv. comm.]{2017MNRAS.465.3558N,2021A&A...653A.151T}, with the blue shaded area showing the lensed models when considering maximum magnifications in the range 10-15 experienced by the background source population as a whole. 
    The models are calculated for the {\it Herschel} SPIRE 500\microns filter, and are identical for the SCUBA-2 450\microns filter \citep[see e.g.,][]{2013MNRAS.432...53G} as the former's bandwidth encompasses that of the latter.
    Despite the extrapolation of over an order of magnitude fainter than the {\it Herschel} lensed galaxy number counts, our inferred source density agrees extremely well with the model predictions.}
    \label{fig:number_counts}
\end{figure*}

\subsection{Inferred Number Density of Strongly-Lensed SMGs}
\label{subsec:inferred-number-counts}

The strongly-lensed SMGs found within the area of the COSMOS field can be used to better predict the number density of these objects across the sky \citep[e.g.][]{2002MNRAS.335L..17C}, and hence how large a population may be visible in the near-infrared by telescopes like {\it Euclid} and {\it Roman}. In addition to this lens, there is another SMG strong lens candidate known to be present in the survey area, albeit without a distinctive arc morphology: ALMA follow-ups of SMGs in the COSMOS field \citep{2022ApJ...929..159C} identify the galaxy AS2COSMOS0002.1 as being strongly lensed. They found the SMG to have a spectroscopic redshift of $z=4.5956 \pm 0.0006$ and a $450$\microns flux density of $22\pm2$\,mJy. They also estimated the magnification to be $\mu=3.0^{+1.4}_{-0.7}$, derived from the presence of several $z\sim1$ foreground galaxies, suggestive of a galaxy group lens.


In Fig.\,\ref{fig:number_counts}, we compare our inferred surface density of strongly gravitationally lensed galaxies with the model predictions presented in \citet{2021A&A...653A.151T}. The lensed models (blue) fit the bright lensed {\it Herschel} number counts from \citet[][dark blue circles]{2017MNRAS.465.3558N} very well; our inferred lensing number count of $27\pm19$ per square degree (dark green diamond) based on the two lenses identified within the {\it JWST}-mapped STUDIES area agrees strikingly well with the model prediction, despite the extrapolation to fainter flux densities of a factor of $\sim\times10$. For comparison we also show the model unlensed number counts (orange line), strongly constrained by the fit to the observed unlensed number counts (red data points). For the \citet[][upward triangles]{2013MNRAS.436.1919C} and \citet[][red circles]{2017MNRAS.464.3369Z} surveys, we report the integral counts in those studies, while for the \citet[][downward triangles]{2013ApJ...776..131C} survey we have recalculated the integral counts using the reported differential number counts, differing slightly from the methodology adopted in \citet{2017MNRAS.465.3558N} and \citet{2021A&A...653A.151T}. We incorporate the asymmetry of the reported uncertainties using the procedure\footnote{Implementation in Python at\newline {\tt https://github.com/anisotropela/add\_asym}} in Appendix B of \citet{2019A&A...627A..84L}. We also show the integral counts implicit from the $P(D)$ analysis of \citet[][red squares]{2010MNRAS.409..109G}, using the same methodology to incorporate asymmetric uncertainties; we base our estimates on their differential number count constraints using a cosmic far-infrared background prior from the Far Infrared Absolute Spectrophotometer (FIRAS). The unlensed model agrees strikingly well with the FIRAS-constrained data and most of the other blank-field survey data, with discrepancies attributable to large-scale structure.

The flux densities for integral number counts are taken to be the faintest end of the faintest bin of the corresponding differential number counts used for that measurement. For our data point we use the median 450\microns flux density of the STUDIES catalogue (Gao et al., in preparation), rather than the minimum SMG flux density or noise limit, as the nature of the SCUBA-2 observations results in the noise limit systematically increasing from the beam centre to the outskirts, resulting in some SMGs with flux densities that reach below the noise levels experienced by many others. Combined with our limited sample size of lenses, we hence use the median flux density, which is also almost identical to the median 3.5$\sigma$ background used as the threshold for SMG identification.

\section{Discussion}
\label{sec:discussion}

Strong lensing models of SMGs have a firm prediction of a few tens of strongly lensed faint SMGs per square degree (Fig.\,\ref{fig:number_counts}). Since SMGs are typically very red at optical/near-infrared wavelengths they are also mainly detected at near-infrared wavelengths. As such, the near-infrared imaging from the upcoming \textit{Euclid} survey will provide one of the best ways of finding these lensing systems. To investigate how many may be observed in the VIS and $H$-band filters of the {\it Euclid}-Wide survey \citep{2011arXiv1110.3193L}, we perform an order-of-magnitude estimation utilising the population of SMGs in \citet{2020ApJ...889...80L}. We restrict the sample to only Very Large Array (VLA) radio-identified SMGs, as their shorter wavelength counterparts (used to distinguish lensed SMGs) can be easily identified thanks to the higher resolution of the VLA beam compared to that of the SCUBA-2 450\microns beam. No cuts are applied to the sources, as the number density of faint SMGs vastly outnumbers that of the bright population. We interpolate their $K$-band and $L$-band AB magnitudes to Euclid $H$-band, and use the $R$-band AB magnitude for the {\it Euclid} VIS filter. Comparing these to the limiting AB magnitudes of {\it Euclid} VIS and NISP $H$-band \citep[26.2 and 24.4 mag, respectively,][]{2022A&A...662A.112E} suggests that only $84\%$ of these faint SMGs will be detectable in {\it Euclid}-Wide VIS, but $99\%$ will be detectable in {\it Euclid}-Wide NISP. For our value of $27\pm19$ strongly lensed faint SMGs per square degree within the 15000 square degree {\it Euclid}-Wide survey, these estimates imply a projection of $62000\pm44000$ NISP-only {\it Euclid}-Wide strong lenses.

Taking an alternative approach also yields a similar end result, albeit with both projections suffering from the large uncertainties incurred by our limited sample of $2\pm\sqrt{2}$ identified lensed SMGs in the STUDIES area. For this approach, we first assume an ``Eyelash'' SED, and sample from submm source counts following \citet{2013MNRAS.432...53G} and a redshift distribution following the model of \citet{2012ApJ...757L..23B}. From this, we find that at 450\microns fluxes $>5$\,mJy, $57\%$ of the galaxies are detectable in NISP $H$-band above its limiting magnitude, while only $40\%$ are detectable in VIS, with these fractions insensitive to the magnification distributions assumed. Again, based on the {\it Euclid}-Wide survey area and our number density, this implies a projection of $70000\pm49000$ NISP-only {\it Euclid}-Wide strong lenses. In summary, the implied huge number of strongly lensed {\it Euclid}-NISP-only galaxies is about as large as the entire strong lensing catalogue predicted for {\it Euclid}-VIS and famously regarded as transformative for strong lensing \citep[e.g.][]{2015ApJ...811...20C}.

Another notable large population of near-infrared-only strong gravitational lenses is the $z>6$ lensed population. For example, \citet{2015ApJ...805...79M} demonstrated that the {\it Euclid}-Wide survey will detect ${\sim}100$ Lyman-break galaxies at redshifts $z>8$, nearly all of which will be strongly gravitationally lensed, but none of which will be detectable in the VIS instrument. These will appear as near-infrared arcs around nearby $z\sim0.5$-2 galaxies, and will therefore be superficially very similar to the lensed SMG population, which enormously outnumber them. The degeneracy between dusty starbursts at redshifts $z<6$ and ultra-high-redshift galaxies in {\it JWST} surveys has already led to downward revisions of redshift estimates \citep[e.g.][]{2023ApJ...943L...9Z}. This may be alleviated with the addition of photometric data from LSST, albeit at lower angular resolution for deblending the foreground lens from the background source. Promising candidates may also have dusty starburst interpretations tested by follow-up submm/mm-wave photometry \citep[e.g.][]{2023ApJ...943L...9Z}. 

Evolved red galaxies are a further potential source of near-infrared-only lensed systems in future dark energy experiments. For example, the population of extremely red objects (EROs, e.g. $R_{\rm Johnson}-K_{\rm Vega}>5$) are approximately evenly split between dusty star-forming galaxies and evolved red galaxies \citep[e.g.][]{Mannucci+02}, with the latter typically having $H_{\rm AB}-K_{\rm AB}<1$ \citep[e.g.][]{Bergstrom+04}. The {\it Euclid} $H_{\rm AB}<24$ limit would therefore be equivalent to a $K$-band survey at least as deep as \citet{Roche+02}, implying surface densities of at least $\sim6000$\,deg$^{-2}$ for each of dusty star forming galaxies and evolved red galaxies. The former is comparable to the surface density of the faintest SMGs in Fig.\,\ref{fig:number_counts} \citep[as expected given the overlap between EROs and SMGs, e.g.][]{Frayer+04} while the latter is clearly a comparably plentiful population of very red background sources for lensing.

\section{Conclusions}
\label{sec:conclusions}

It is well established that the brightest sub-millimetre galaxies (SMGs) at 450-500\microns wavelengths are almost exclusively strong gravitational lenses, with such galaxies expected to appear very red in the near-infrared. Making use of this, in this work we have searched public {\it JWST} NIRCam and MIRI imaging of the COSMOS field for counterparts of SCUBA-2 450\microns sources in the STUDIES catalogue and identified a strong gravitational lensing system. Modelling of the lens indicates a mass distribution with Einstein radius $\theta_{E}=0.63$ and axis ratio $q=0.77$ that distorts the highly elliptical source into an arc with a magnification of $\mu=3.0$. Performing SED fitting using public photometry along with the known spectroscopic redshift of the foreground lens ($z_{\rm l}=0.360$), the background source was found to be an SMG with a photometric redshift of $z_{s}=3.4\pm0.4$.

This, along with a previously discovered SMG lens in the survey area, supports the prediction of around ten strongly lensed faint SMGs per square degree, which in turn indicates a very large population of tens of thousands of strong gravitational lenses undetectable in the optical that would be prime targets for future dark energy surveys that can image in the near-infrared, such as the {\it Euclid} and {\it Roman} wide-field surveys.

\section*{Acknowledgements}

We thank the anonymous referee for their helpful comments and suggestions. We are grateful to Mattia Negrello for very kindly supplying machine-readable versions of his number count models from \citet{2017MNRAS.465.3558N} and \citet{2021A&A...653A.151T} for Fig.\,\ref{fig:number_counts}.
{\it Herschel} is a European Space Agency (ESA) space observatory with science instruments provided by European-led Principal Investigator consortia and with important participation from the National Aeronautics and Space Administration (NASA). 
The James Clerk Maxwell Telescope is operated by the East Asian Observatory on behalf of The National Astronomical Observatory of Japan; Academia Sinica Institute of Astronomy and Astrophysics; the Korea Astronomy and Space Science Institute; the National Astronomical Research Institute of Thailand; Center for Astronomical Mega-Science (as well as the National Key R\&D Program of China with No. 2017YFA0402700). Additional funding support is provided by the Science and Technology Facilities Council of the United Kingdom and participating universities and organisations in the United Kingdom and Canada. 
Additional funds for the construction of the Submillimetre Common-User Bolometer Array 2 (SCUBA-2) were provided by the Canada Foundation for Innovation. 
The submillimeter observations used in this work include the SCUBA-2 COSMOS program (S2COSMOS; program code M16AL002), SCUBA-2 Ultra Deep Imaging EAO Survey program (STUDIES; program code M16AL006), SCUBA-2 Cosmology Legacy Survey program (S2CLS; program code MJLSC01) and the PI program of \citet[][program codes M11BH11A, M12AH11A, and M12BH21A]{2013MNRAS.436.1919C}.
JP and SS were partly supported by the ESCAPE project; ESCAPE – The European Science Cluster of Astronomy \& Particle Physics ESFRI Research Infrastructures has received funding from the European Union’s Horizon 2020 research and innovation programme under Grant Agreement no. 824064. LCH was supported by the National Science Foundation of China (11721303, 11991052, 12011540375, 12233001), the National Key R\&D Program of China (2022YFF0503401), and the China Manned Space Project (CMS-CSST-2021-A04, CMS-CSST-2021-A06). HS acknowledges the support from the National Research Foundation of Korea grant No.2021R1A2C4002725 and No.2022R1A4A3031306, funded by the Korea government (MSIT).  C.-C.C. acknowledges support from the National Science and Technology Council of Taiwan (NSTC 109-2112-M-001-016-MY3 and 111-2112M-001-045-MY3), as well as Academia Sinica through the Career Development Award (AS-CDA-112-M02). HSH acknowledges the support by the National Research Foundation of Korea (NRF) grant funded by the Korea government (MSIT) (No. 2021R1A2C1094577). LF gratefully acknowledges the support of the National Natural Science Foundation of China (NSFC, grant No. 12173037, 12233008), the CAS Project for Young Scientists in Basic Research (No. YSBR-092). MJM acknowledges the support of the National Science Centre, Poland through the SONATA BIS grant 2018/30/E/ST9/00208 and the Polish National Agency for Academic Exchange Bekker grant BPN/BEK/2022/1/00110.

This work made use of \textsc{Astropy}:\footnote{\tt http://www.astropy.org} a community-developed core Python package and an ecosystem of tools and resources for astronomy \citep{2013A&A...558A..33A,2018AJ....156..123A,2022ApJ...935..167A}. Data analysis made use of the Python packages \textsc{NumPy} \citep{2020Natur.585..357H} and \textsc{SciPy} \citep{2020SciPy-NMeth}. Some figures were made with the Python package \textsc{matplotlib} \citep{Hunter:2007}. SED fitting involved the Flexible Stellar Population Synthesis for Python package \textsc{pythonFSPS} \citep{2009ApJ...699..486C,2010ApJ...712..833C}, and lens modelling was performed using the Python package \textsc{lenstronomy} \citep{2018PDU....22..189B,2021JOSS....6.3283B}.

\section*{Data Availability}

The {\it Herschel} SPIRE data can be downloaded at {\tt https://www.h-atlas.org}, while reduced and calibrated science-ready ALMA data are available on the ALMA Science Archive at {\tt https://almascience.eso.org/asax/}\,. The JCMT Science Archive at the Canadian Astronomical Data Centre, {\tt http://www.cadc-ccda.hia-iha.nrc-cnrc.gc.ca/en/jcmt/}\,, contains all raw and processed data from the JCMT's current instrumentation. The COSMOS-Web and PRIMER calibration level 3 data, along with other publicly available {\it JWST} data, are available through the Space Telescope Science Institute (STScI) Mikulski Archive for Space Telescopes (MAST) Portal\footnote{\tt https://mast.stsci.edu/portal/Mashup/Clients/Mast/Portal.html}.



\bibliographystyle{mnras}
\bibliography{bibliography}







\bsp	
\label{lastpage}
\end{document}